# Deposition and Visualization of DNA Molecules on Graphene That is Obtained with the Aid of Mechanical Splitting on a Substrate with an Epoxy Sublayer

A. V. Frolov[a, b], N. A. Barinov[c], D. V. Klinov[c], V. V. Koledov[a], P. V. Lega[a],
A. P. Orlov[a, d], and A. M. Smolovich[a, *]

[a]Kotel′nikov Institute of Radio Engineering and Electronics, Russian Academy of Sciences, Moscow, 125009 Russia
[b]Moscow Institute of Physics and Technology, Dolgoprudnyi, Moscow oblast, 141700 Russia
[c]Federal Scientific Clinical Center of Physicochemical Medicine, Moscow, 119435 Russia
[d]Institute of Nanotechnology of Microelectronics, Russian Academy of Sciences, Moscow, 115487 Russia
*e-mail: asmolovich@petersmol.ru

**Abstract**—Controlled deposition of DNA on graphene films obtained with the aid of mechanical splitting of graphite on a substrate with an epoxy sublayer is demonstrated. The DNA molecules are visualized using AFM.

## INTRODUCTION

Considerable recent interest in the DNA–graphene hybrid structures [1–5] has been driven mainly by the applications of such structures in biosensors, in particular, medical diagnostics. Note also that graphene can be used as a carrier for DNA storage, hybridization, targeted assembling, and interaction of complementary chains [6]. It is of interest to study graphene structuring with the aid of deposition of metallized DNA that can be considered as molecular wire [7, 8]. The deposition may lead to variations in the conductivity of graphene and DNA molecules. Complicate technology of the DNA deposition on graphene depends on the method for fabrication of graphene.

A method for fabrication of graphene based on mechanical splitting of graphite using epoxy adhesive has recently been proposed in [9, 10]. On the one hand, the method is simple in comparison with the widely spread methods for growth of graphene films [11] or methods based on plasmachemical etching [12, 13]. On the other hand, the method makes it possible to increase the film area in comparison with alternative methods for mechanical splitting, in particular, the Geim–Novoselov method [14]. However, the main advantage of the method lies in reliable fabrication of almost perfect freshly cleaved graphene surface. The method has been used to obtain single-layer graphene films with an area of up to $200 \times 200$ μm$^2$ and atomic-thin graphite with an area of up to $1000 \times 1000$ μm$^2$. Quantum effects revealed in the structures based on such films indicate relatively high quality of the films and original crystal [15–19]. We assume that 2D graphene crystals fabricated with the aid of mechanical splitting of graphite using epoxy adhesive may serve as sensitive elements of biosensors. The purpose of this work is the development of technology for deposition of DNA molecules on graphene samples fabricated using the new method.

## 1. EXPERIMENTAL PROCEDURE

The method for fabrication of graphene of [9, 10] lies in thinning of bulk graphite crystals using adhesion tape and can be considered as a modification of the method of [14]. Natural graphite serve as original crystals, since such a material contains relatively large single crystals, forms atomic-smooth surface upon exfoliation, and, in accordance with the results of [20, 21], is superior (with respect to quality) to highly oriented pyrolytic graphite (HOPG) that has been used in [14]. Upper layers of the crystal are removed with the aid of adhesion tape to obtain mirror-smooth regions with an area of up to 1 mm$^2$. Then, the surface of the graphite crystal is attached to the substrate using epoxy adhesive and thinned using adhesion tape to nanometer-scale thicknesses. Thus, we obtain atomic-thin layered single crystals of graphite including single-layer graphene attached to the substrate by the adhesive agent (Fig. 1). Two methods are used to determine the quantity of the graphene layers. In the first method, we measure the optical transmittance of the sample in the visible spectral range and perform calculations using the fact that each graphene layer provides absorbance of about 2.3% [22]. For such measurements, the sample must be placed on a transparent (e.g., glass) substrate. In the second method, the number of graphene layers is determined with the aid of Raman scattering [23].

For the deposition of DNA molecules on graphene, we use a procedure that is similar to the procedure for deposition of DNA molecules on graphite [24–27]. Solution of graphite modifier (GM) (($CH_2$)$_n$($NCH_2CO$)$_m$–$NH_2$ from Nanotyuning, Russia) with a concentration of 0.1% and a volume of 100 μL is deposited on the graphene surface. After 1-min-long exposure, the GM is removed using nitrogen jet and the substrate is dried. Molecules of duplex DNA from Escherichia virus Lambda with a concentration of 1 μg/mL are deposited from the solution containing 10 mM Tris-HCl (pH 7.6) and 1 mM EDTA on the surface of the modifier for 1 min, and, then, the drop with dissolved DNA is removed using the nitrogen jet. The monolayer of the modifier represents lamellar structures that are epitaxially crystallized by intermolecular H bonds on the graphene surface. The deposition of the modifier is needed for attachment of unfolded DNA molecules to the surface, since DNA molecules weakly interact with pure graphene and exhibit twisting, folding, and shifting upon passage of the droplet meniscus. When the droplet is dried rather than removed using the nitrogen jet, the impurities contained in the solution at mass fractions of greater than $10^{-6}$ are precipitated as a rough layer with a thickness of about 1 nm that is comparable with the DNA thickness. This circumstance impedes the DNA identification using AFM.

We employ AFM to monitor the DNA deposition. The landscapes of samples are measured in the semi-contact resonance regime on an NT-MDT Integra Prima setup using NOVA 1.1 software. High-resolution supersharp silicon cantilevers from Nanotyuning are used for the measurements. The resonance frequencies of the cantilevers range from 190 to 325 kHz, the radius of curvature of the tip is less than 2 nm, and the apex angle is less than 22°. The amplitude of the free oscillations of the cantilever in air ranges from 1 to 1o nm. The automatically maintained amplitude of the cantilever oscillations in the vicinity of the surface (SetPoint parameter) is fixed at a level of 60–70% of the amplitude of the cantilever oscillations in air.

The NOVA 1.1 software from NT-MDT is used for signal processing, digitization, and imaging. The experimental results are represented as 2D images in which light and dark surface fragments correspond to hills and wells, respectively (Figs. 2 and 3). The NOVA Image Analysis 2.0 software is used for data processing.

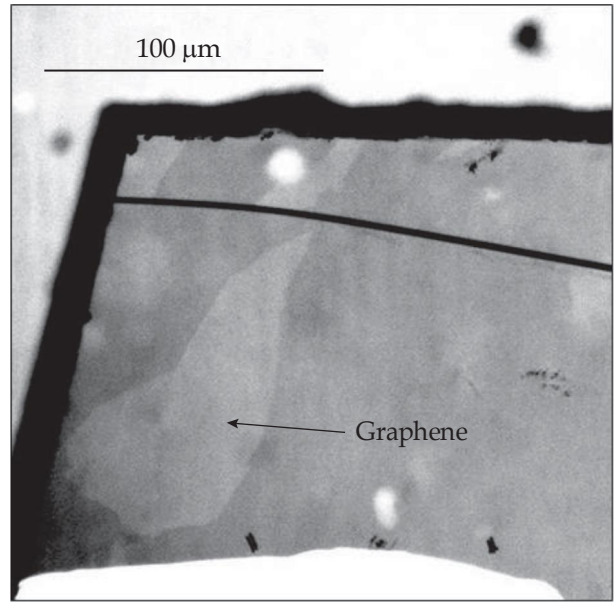

**Fig. 1.** Images of graphene flakes obtained with the aid of optical microscopy in the reflection mode.

## 2. RESULTS AND DISCUSSION

Figure 2 presents the AFM images of DNA molecules on the graphene surface. It is seen that the proposed method makes it possible to obtain relatively uniform distribution of DNA on the graphene surface. DNA molecules form relatively strong bonds with modified graphene, so that even small modifier-free fragments are covered with DNA molecules. The measurements of heights in such fragments (dark areas in Fig. 2a) are used to estimate the thickness of the modifier layer (about 0.7 nm). DNA molecules can also be deposited on interfaces with different numbers of the graphene layers (like the graphene-bigraphene composition) (Fig. 2b). DNA molecules are not detected on the surface of the epoxy adhesive that is not covered with graphite or graphene (upper right-hand region in Fig. 2b).

Practical results show that the modifier must be deposited on graphene immediately after splitting, since relatively long exposure of the graphene surface to air leads to contamination, so that the modifier cannot be deposited after short exposures. The graphite sur- face with the modifier is more stable in air: DNA was successfully deposited on the graphene sample that was stored under forvacuum conditions over three months after deposition of the modifier (Fig. 2c). For such a sample, the DNA molecules are stretched along the direction of the gas jet owing to relatively weak interaction with the modifier. For better adhesion of DNA to modifier, the droplet with dissolved DNA is kept on the sample over a longer time interval (2– 3 min). However, in this case, the deposition of DNA is accompanied by precipitation of impurities: balls

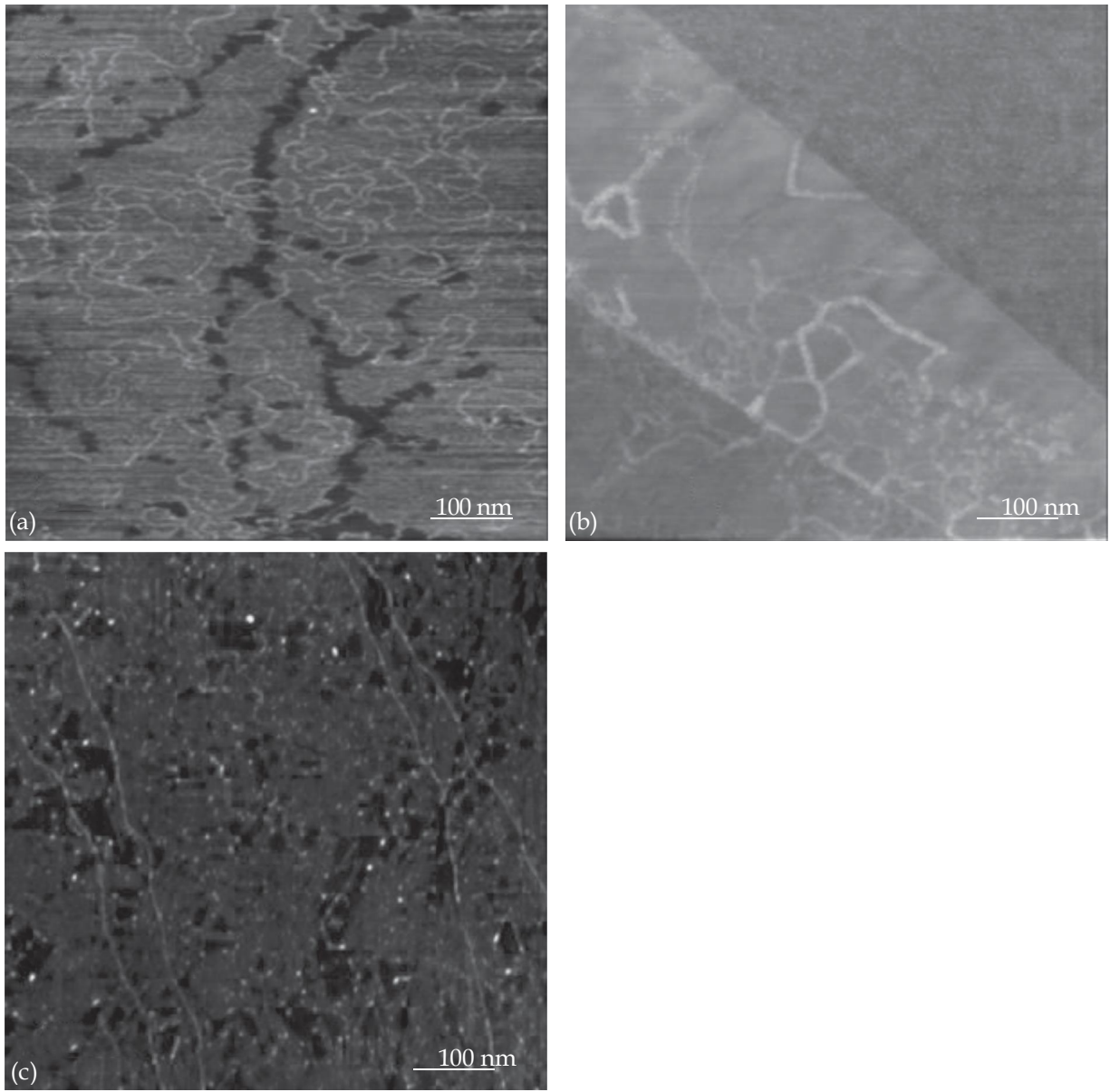

**Fig. 2.** AFM images of DNA molecules on (a) graphene surface with the modifier layer, (b) graphene–bigraphene–substrate interface with the modifier layer, and (c) graphene samples stored under forvacuuum conditions after modifier deposition.

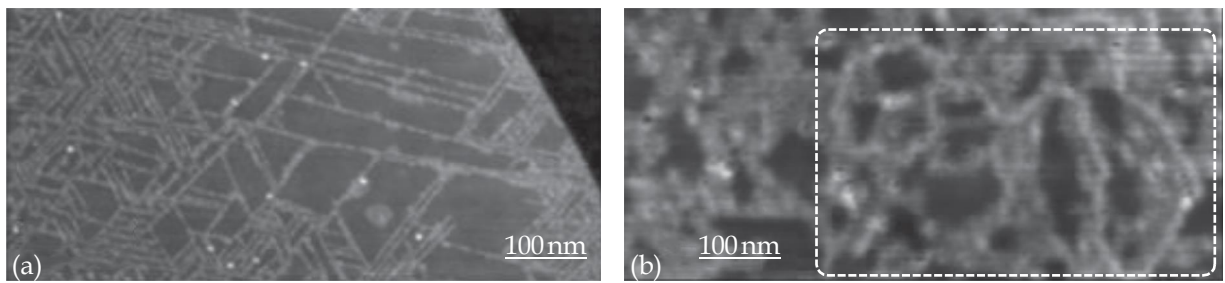

**Fig. 3.** AFM images of (a) graphene fragments formed on the Si/SiO$_2$ substrate using the Geim–Novoselov method [11] with the deposited modifier (the presence of noninterfaced lamellas of modifier indicates surface contamination of graphene) and (b) single DNA molecules (in the frame).

with a diameter of several nanometers are observed on the graphene surface (Fig. 2c).

For comparison of the efficiencies of the proposed and conventional methods, we make attempts at deposition of the modifier and DNA molecules on graphene fabricated using the classical method of [14] on the substrate made of heavily doped silicon with 300-nm-thick layer of $SiO_2$. Problems emerge at the stage of the modifier deposition: the modifier poorly covers the graphene surface being deposited as interfaced lamellas oriented along the crystalline directions of graphene (Fig. 3a). The presence of adhesive agent from the adhesive tape on the substrate is a possible reason for such problems. We have revealed the following regularity: the probability with which the modifier is well deposited and DNA molecules are found on the surface of the modifier increases with an increase in the distance from the flake to the area covered with adhesive agent. At such fragments that are far from the regions covered with the adhesive agent, we find single DNA molecules deposited on graphene (Fig. 3b). Dissolution of the adhesive agent does not lead to better results, since a thin layer of contaminants that survives on the graphene surface prevents the modifier deposition.

We have also tried to measure the transport characteristics of graphene prior to and after DNA deposition. In our opinion, the observed variations in the graphene conductivity after the deposition predominantly result from variation in the position of the Fermi level relative to the Dirac point [28]. The problem lies in the fact that the graphene surface is extremely sensitive to the interaction with environment, so that the Fermi level can be shifted due to both DNA deposition and interaction of the graphene surface with air and precipitated impurities (Fig. 2c).

## CONCLUSIONS

We have proposed laboratory technology for deposition of DNA molecules on graphene that is obtained with the aid of mechanical splitting of natural graphite using epoxy adhesive. The procedure involves preliminary deposition of the modifier layer on the freshly prepared graphene. DNA molecules are deposited on top of the modifier layer. The stages of the procedure are monitored using AFM. The comparison have shown that the proposed procedure for the deposition of DNA molecules is superior (with respect to quality and technological simplicity) to alternative methods. The results of this work can be used in the development of nanoelectronic sensors for biomedical applications.

## ACKNOWLEDGMENTS

This work was supported by the Russian Science Foundation (project no. 17-19-01748).

*Translated by A. Chikishev*